\documentclass[aps,pra,twocolumn]{revtex4-1}

\pdfoutput=1

\usepackage{amsfonts,amsmath,amssymb}
\usepackage{mathrsfs}
\usepackage{nicefrac}
\usepackage{graphicx}
\graphicspath{{graphics/}} %Indica la cartella con le immagini
\usepackage{color}
\usepackage{soul} 
\usepackage{hyperref}
\usepackage{accents}
\usepackage[normalem]{ulem}
\definecolor{Teal}{RGB}{45,48,146}
\usepackage{cleveref}
\hypersetup{
	citecolor=Teal,
    colorlinks=true, 
    linktoc=page,    %set to all if you want both sections and subsections linked 
    linkcolor=black,  
    urlcolor=black
}

\usepackage{bookmark}
\bookmarksetup{
  numbered, 
}
\usepackage{lmodern} % math, rm, ss, tt
\usepackage[T1]{fontenc}
\usepackage[utf8]{inputenc} %per scrivere le lettere accentate direttamente da tastiera
\usepackage{amssymb,amsmath} %simboli
\usepackage{textcomp}
\usepackage{mathrsfs}
\usepackage{braket}
\usepackage{slashed}
\usepackage{tensor}
\usepackage[subnum]{cases}
\usepackage{mathtools}

\newcommand{\de}{\partial}

\newcommand{\del}{\nabla}

\newcommand{\T}{\Theta}
\newcommand{\w}{\omega}

\newcommand{\vc}[1]{\mathbf{#1}}

\newcommand{\h}{\hbar}

\newcommand{\tld}[1]{\widetilde{#1}}

\begin{document}

\title{Ergoregion instabilities in rotating two-dimensional Bose--Einstein condensates:\\new perspectives on the stability of quantized vortices}

\author{Luca Giacomelli}
\email[]{luca.giacomelli-1@unitn.it}
\author{Iacopo Carusotto}
\email[]{iacopo.carusotto@unitn.it}
\affiliation{INO-CNR BEC Center and Dipartimento di Fisica, Universit\`a di Trento, via Sommarive 14, I-38050 Povo, Trento, Italy}

\begin{abstract}
	We investigate the stability of vortices in two-dimensional Bose--Einstein condensates. In analogy with rotating spacetimes and with a careful account of boundary conditions, we show that the dynamical instability of multiply quantized vortices in trapped condensates persists in untrapped, spatially homogeneous geometries and has an ergoregion nature with some modification due to the peculiar dispersion of Bogoliubov sound. Our results open new perspectives to the physics of vortices in trapped condensates, where multiply quantized vortices can be stabilized by interference effects and singly charged vortices can become unstable in suitably designed trap potentials. We show how superradiant scattering can be observed also in the short-time dynamics of dynamically unstable systems, providing an alternative point of view on dynamical (in)stability phenomena in spatially finite systems.
\end{abstract}

\maketitle

%%%%%%%%%%%%%%%%%%%%%%%%%%%%%%%%%%%%%%%%%%%%%%%%%%%%%%%%%%%%%%%%%%%%%%%%%%%%%%%%%%%%%%%%%%%%%

\section{Introduction}
Quantized vortices are one of the key features of superfluids and Bose--Einstein condensates (BECs) and have received a great deal of attention in the last decades, both theoretically and experimentally~\cite{fetter2009rotating,pitaevskii2016bose}. The studies of such configurations began with the very introduction of the Gross--Pitaevskii equation in the early 1960's in~\cite{pitaevskii1961vortex}, where singly quantized vortices in an infinite condensate were considered, and received a further boost with the experimental realization of atomic BECs in optical and/or magnetic traps.
In this context, the focus is naturally on trapped geometries~\cite{dodd1997excitation,rokhsar1997vortex,pu1999coherent,svidzinsky2000stability,mottonen2003splitting,shin2004dynamical,lundh2006dynamic,takeuchi2018doubly}, while much less attention is typically paid to spatially infinite geometries. Only quite recently a convincing evidence that doubly quantized vortex are unstable in a spatially uniform BEC was in fact reported~\cite{takeuchi2018doubly}, while dynamical stability of such configurations had been previously claimed by several authors~\cite{aranson1996stability,svidzinsky2000stability}. More generally, a full understanding of the microscopic mechanisms determining the stability of quantized vortices in generic geometries is still missing.

An interesting way to look at this paradigmatic problem of condensed matter physics is the one offered by Analogue Gravity~\cite{barcelo2011analogue}, that relies on the fact that collective excitations modes on top of a moving medium are described by the same equations as a massless scalar field on a curved spacetime. In particular, vortices in various fluids have been often proposed as analogues of rotating spacetimes, since they can display key ingredients such as an \textit{horizon} and an \textit{ergoregion}. In this context, a particular attention has been given to the phenomenon of \textit{rotational superradiance}~\cite{brito2015superradiance}, that is the amplified scattering of radiation from a rotating black hole.

This phenomenon relies on an {\em energetic instability} of rotating black holes that can become a {\em dynamical} one under suitable boundary conditions. Amplified superradiant scattering from a dynamically stable system is found if perturbations are free to propagate away both in the asymptotic outward direction and in the inward one across the horizon, as it happens for massless fields around rotating Kerr black holes in an asymptotically flat spacetime. On the other hand, superradiance can get self-stimulated in the presence of sizable reflections on either side, which give rise to radiation modes with complex frequencies and, thus, to exponentially growing perturbations in time. If reflection occurs on the inner side, e.g. in a rotating spacetime displaying an ergoregion but no horizon, the ensuing dynamical instability is called {\em ergoregion instability} \cite{friedman1978ergosphere, comins1978ergoregion}. If reflection occurs in the external space away from the black hole, a {\em black hole bomb} is instead found, e.g. for rotating black holes in a AdS background or massive fields around Kerr black holes~\cite{hawking1999charged,cardoso2004black}.

In the analogue systems context, superradiant scattering has been widely studied from the theoretical point of view~\cite{basak2003superresonance,slatyer2005superradiant,federici2006superradiance,richartz2015rotating,cardoso2016detecting} and has been recently observed in experiments using surface gravity waves on top of a water flow configuration displaying a draining vortex \cite{torres2017rotational}. Ergoregion instabilities have also been theoretically studied in purely rotating vortex configurations in ideal fluids~\cite{oliveira2014ergoregion, oliveira2018ergoregion}. We are here interested in quantized vortices in superfluids: their purely azimuthal $v_\theta\propto 1/r$ irrotational flow pattern becomes supersonic in the vicinity of the vortex core and corresponds to an analogue rotating spacetime with an ergoregion but no horizon. This naturally suggests the possibility of observing ergoregion instabilities in infinite condensates. But given that outer reflections are also naturally present in trapped geometries, it is also tempting to connect the known instabilities of multiply charged vortices in harmonically trapped BECs to black-hole-bomb-like instabilities. Unraveling the different instability mechanisms that are at play in different geometries is the main subject of this work.

Together with the intriguing links with the space-time instabilities, a specific attention must be paid also to the microscopic features of the atomic condensate. The sudden density drop in the vortex core and the superluminal dispersion of collective excitations at high momenta prevent a straightforward use of the hydrodynamic approximation that is at the basis of the gravitational analogy. 
In this work, we carry out a microscopic study of the Bogoliubov collective excitations around vortices of different charges in spatially homogeneous two-dimensional BECs beyond the hydrodynamic approximation. With a careful consideration of the boundary conditions, we confirm the conclusion that singly quantized vortices are stable whereas doubly quantized ones are made dynamically unstable by an intrinsic instability of the vortex core, analogous to an ergoregion instability.

We then extend our analysis to vortices with higher charge to further characterize the instabilities in the different angular momentum channels: additional similarities with ergoregion instabilities are found. From this we  conclude that the dynamical instabilities of multiply quantized vortices are condensed matter analogues of the ergoregion instabilities of rotating spacetimes, with some modifications due to dispersive effects. Our results also shine new light on the known results for trapped BECs, showing that the instability is not induced by the trap and, thus, is not related to black-hole-bomb type mechanisms. The only effect of the trap is rather to modulate the instability rate via interference mechanisms and even suppress it in specific regimes. 
Application of our formalism to singly quantized vortices brings the unexpected consequence that their celebrated dynamical stability is not a general fact, but a consequence of the spatially homogeneous or harmonic trap geometries usually considered in the literature: more complex configurations showing an inner density bump followed by a constant density plateau turn out to be dynamically unstable against the vortex spiralling out even at zero temperature. As a final point, we discuss how superradiant scattering interplays with the dynamical instability and we highlight how it can be observed on short time scales also in the presence of dynamical instabilities.

The structure of the article is the following. In Sec.\ref{sec:basics} we review the general formalism and we present the system under consideration. In Sec.\ref{sec:known} we recap the state-of-the-art in the theory of the dynamical stability of vortices in trapped condensates. In Sec.\ref{sec:charge-2} we investigate the dynamical instability of a doubly quantized vortex in a spatially homogeneous condensate and we highlight the subtle issues related to the boundary conditions. In Sec.\ref{sec:higher} we present our results on the stability of higher charge vortices and we draw connections to the purely hydrodynamic theory underlying the gravitational analogy. In Sec.\ref{sec:charge1} we present our unexpected results on the stability of singly-charged vortices. The interplay between dynamical instabilities and superradiant scattering is discussed in Sec.\ref{sec:superradiance}. Conclusions and future perspectives are finally presented in Sec.\ref{sec:conclu}.

\section{Vortices and the linear problem}
\label{sec:basics}
Dilute Bose-Einstein condensates are well described at the mean field level by the Gross--Pitaevskii equation (GPE) for a scalar classical field describing the order parameter $\Psi(\vc{r},t)$ \cite{pitaevskii2016bose},
\begin{equation} 
	i\h\de_t \Psi = \left[-\frac{\h^2\del^2}{2M}+g|\Psi|^2+V_{\mathrm{ext}}\right]\Psi,
\label{eq:GPE} 
\end{equation}
where $g$ is the interparticle interaction constant, $M$ is the atomic mass and $V_{\mathrm{ext}}(\mathbf{r})$ is an external trapping potential. We consider the case of two spatial dimensions, which simplifies the treatment and is a good approximation for a pancake-shaped condensate tightly confined in the third direction. 

This equation can be rewritten in the density-phase representation of the order parameter $\Psi(\vc{r})=\sqrt{n(\vc{r})}\,e^{i\Theta(\vc{r})}$ as:
\begin{equation}\label{eq:GPE-densityphase}
	\begin{cases}\displaystyle
		\de_t n = -\del\cdot\left(n\frac{\h\del\T}{M}\right)\\
		\displaystyle
		\h\de_t \T + \frac{\left(\h\del\T\right)^2}{2M}+gn+V-\frac{\h^2}{2M}\frac{\del^2\sqrt{n}}{\sqrt{n}}=0.
	\end{cases}
\end{equation}
These are the usual continuity and Euler equations for a perfect fluid, except for the last term in the second equation that is called \textit{quantum pressure} and stems from the kinetic energy of quantum particles. This term can be neglected when density variations happen over distances much larger than the so-called healing length $\xi=\hbar/(mgn)^{1/2}$: this regime is usually called the \textit{hydrodynamic regime}. 

Vortices located at the center of a cylindrically-symmetric system are stationary solutions of equation \eqref{eq:GPE} of the form 
\begin{equation}
	\Psi_\ell(\vc{r},t)=f(r)e^{i\ell\theta}e^{-i\mu t/\h},
\end{equation}
where $\mu$ is the chemical potential of the condensate and $\ell$ is a number that, for the order parameter to be single valued, must be an integer. This expresses the \textit{quantization of the circulation} of the vortex and we refer to $\ell$ as the \textit{charge} of the vortex. 
Vortices of charge $\ell$ have a purely azimuthal velocity profile of the form
\begin{equation}
	v_\theta(r)=\frac{\h}{M}\frac{\ell}{r},
\end{equation}
which turns supersonic in the vicinity of the vortex core.

To avoid singularities at $r=0$, the (real-valued) amplitude $f(r)$ of the order parameter must go to zero for $r\to0$ so that in the center of vortex the density is depleted. For a vortex in an otherwise infinite and spatially homogeneous condensate the order parameter tends to a constant $f(r)\to f_\infty$ at large distances, while the chemical potential is $\mu=gn_\infty=g|f_\infty|^2$. For a vortex of charge $\ell$, the current becomes supersonic at a radius $r_E\sim\ell\xi$, where the healing length $\xi=\hbar/(mgf_\infty^2)^{1/2}$ is calculated for the asymptotic value of the density. The healing length can also be seen as the distance at which the density reaches approximately half of the value at infinity \cite{pitaevskii2016bose}. Near the core of the vortex we are hence away from the hydrodynamic regime.

A standard tool to study the stability of vortices is the so-called Bogoliubov approach, where one linearizes the GPE \eqref{eq:GPE-densityphase} around its stationary state $\Psi_\ell$ and looks for the eigenmodes of the linearized dynamics~\cite{castin2001bose}. To this purpose, we consider the (small) perturbation $\delta\Psi(\vc{r},t)$ and its complex conjugate $\delta\Psi^*$ as independent variables and group them into the spinor $(\delta\Psi,\ \delta\Psi^*)^T$. 

In the cylindrically symmetric geometries under consideration in this work, we can decompose the perturbation in its angular momentum $m$ components
\begin{equation}
	\begin{pmatrix}
		\delta\Psi\\ \delta\Psi^*
	\end{pmatrix}(r,\theta,t)
	= e^{im\theta}
	\begin{pmatrix}
		e^{i\ell\theta}e^{-i\mu t/\h} u_\phi(r,t) \\ 
		e^{-i\ell\theta}e^{i\mu t/\h} v_\phi(r,t) 
	\end{pmatrix}.
\end{equation}
and focus our attention on the radial dynamics. For each $m$ component, the time evolution for the radial spinor $\ket{\phi}(r,t):=(u_\phi(r,t),v_\phi(r,t))^T$ is given by the Bogoliubov--de Gennes (BdG) equation 
\begin{equation}
\label{eq:bogodegennes}
	i\hbar\de_t \ket{\phi}=\mathscr{L}_{\ell,m} \ket{\phi}
\end{equation}
with the radial Bogoliubov operator
\begin{equation} 
	\mathscr{L}_{\ell,m} = 
	\begin{bmatrix}
		D_+ + V_{\mathrm{ext}} + 2gf^2-\mu & gf^2\\
		-gf^2 & -(D_- + V_{\mathrm{ext}} + 2gf^2-\mu)
	\end{bmatrix}
	\label{eq:BdGmatrix}
\end{equation}
and
\begin{equation}
	D_\pm = \frac{\h^2}{2M}\left(-\de_r^2-\frac{\de_r}{r}+\frac{(\ell\pm m)^2}{r^2}\right).
\end{equation}

In contrast to the unitary evolution of the usual Schr\"odinger equation, the evolution (\ref{eq:bogodegennes}) is $\sigma_3$-pseudo-unitary, i.e. $\sigma_3\mathscr{L}_{\ell,m}^\dag\sigma_3=\mathscr{L}_{\ell,m}$, so that the conserved inner product
\begin{equation}\label{eq:bogo-norm}
	\bra{\psi}\sigma_3\ket{\psi}=2\pi\int \mathrm{d}r\;r\left[ u^*_\psi(r)u_\psi(r) - v^*_\psi(r)v_\psi(r) \right]
\end{equation}
is non-positive definite; the associated norm can hence also be negative and zero. The energy of an eigenmode $\ket{\psi_i}$ of the BdG matrix (\ref{eq:BdGmatrix}) with frequency $\w_i$ is given by 
\begin{equation}
 E_i=\bra{\psi_i}\sigma_3\ket{\psi_i}\h\w_i,
 \label{eq:norm}
\end{equation}
so that, for example, negative-norm (positive-norm) modes with a positive (negative) frequency have negative energy and zero-energy modes can also exist.

Since the matrix $\mathscr{L}_{\ell,m}$ is not hermitian, complex eigenvalues are also possible, corresponding to modes that exponentially decay or grow in time. The symmetries of the BdG matrix impose that, if $\w_i$ is an eigenfrequency, also $\w_i^*$ is a (possibly different) eigenfrequency and that $(\w_i-\w_j^*)\bra{\psi_j}\sigma_3\ket{\psi_i}=0$ for a pair of generic $i,j$ eigenmodes. As a consequence, complex-frequency modes come in \textit{pseudo-degenerate} (sharing the same real part) pairs of decaying and dynamically unstable zero-norm modes. According to \eqref{eq:norm}, zero norm implies zero energy: in physical terms the exponential growth of unstable modes corresponds to the simultaneous creation of particles and antiparticles with opposite energies, which leaves the total energy unchanged.

Another property of the BdG matrix is the so-called \textit{particle-hole symmetry}, that in our cylindrically symmetric geometry is expressed by the fact that the spectrum at $-m$ is specular to the one at $m$. In detail, there exist pairs $i,j$ of eigenvectors and eigenvalues at $\pm m$ that are related to each other by
\begin{equation}
	\begin{pmatrix}
		u_{-m,j}\\ v_{-m,j}
	\end{pmatrix}
	=
	\begin{pmatrix}
		v_{m,i}\\ u_{m,i}
	\end{pmatrix}
	;\hspace{.5cm} \w_{-m,j}=-\w_{m,i},
\end{equation}
so that both the sign of the frequency and the norm of the mode are inverted. In what follows, we can thus restrict our attention to positive values of $m$ only with no loss of information.

In what follows, it will be useful to consider also a different shape of the radial BdG equations. This form can be obtained for example by linearizing equation \eqref{eq:GPE-densityphase}, written in terms of density and phase perturbations $\delta\tld n:=\delta n/n$ and $\delta\T$ 
\begin{equation}\label{eq:bdg-densphase}
		\de_t
	\begin{pmatrix}
		\delta\Theta\\ \delta\tld n
	\end{pmatrix}
	=
	\begin{bmatrix}
		\displaystyle -i\frac{\h}{M}\frac{\ell m}{r^2} & & \displaystyle -\frac{\tld D}{2} - \frac{Mc_s^2}{\h}\\
		\\
		\displaystyle 2\tld D & & \displaystyle -i\frac{\h}{M}\frac{\ell m}{r^2} 
	\end{bmatrix}
	\begin{pmatrix}
		\delta\Theta\\ \delta\tld n
	\end{pmatrix}
\end{equation}
with
\begin{equation}
	\tld D= \frac{\h}{2M}\left(-\de_r^2 - \frac{\de_r}{r} -\frac{\de_r f^2}{f^2}\de_r +\frac{m^2}{r^2}\right).
\end{equation}
For slowly spatially varying (\textit{long-wavelength}) perturbations the derivative term in the upper right element is negligible. The resulting equation has the shape of a Klein--Gordon (KG) equation in a curved spacetime with the local speed of sound $c_s=\sqrt{g f^2/m}$ playing the role of the speed of light (see for example \cite{fulling1989aspects} for this shape of the KG equation).

This is the core idea of analogue gravity \cite{barcelo2011analogue}, where the curved spacetime is determined by the flow profile of the condensate. For instance, vortex solutions in an infinite BEC correspond to asymptotically flat rotating spacetimes with an \textit{ergoregion} corresponding to the region where $v_\theta$ is supersonic. As usual, ergoregions are characterized by the existence of negative energy modes for the field with respect to the asymptotic definition of energy; in the analogue these negative energy modes are associated to the energetic instability of regions of supersonic flow, where superfluidity may break down~\cite{pitaevskii2016bose,castin2001bose}.

The spectral properties of the BdG equations presented above are shared by the KG equation, to which, as we said, the BdG equation reduces in the long-wavelength limit. At larger wavevectors above the inverse healing length, the dispersion relations for the two problems are however different: while for the KG equation it remains linear at all wavevectors, for the BdG equation it turns quadratic above $\xi^{-1}$, which corresponds to a Lorentz invariance breaking at small scales. In spite of these differences, it is useful to look at the problem of the vortex  stability starting from what is known for gravitational systems and look for possible modifications due to the deviations from the linear dispersion of the hydrodynamic limit and to the microscopic structure of the vortex core.

Notice that, while the hydrodynamic approximation on \eqref{eq:GPE-densityphase} automatically implies taking the long-wavelength limit for the fluctuations, leading to the non-dispersive KG problem, the long-wavelength approximation for the perturbation field can formally be performed even for background density profiles outside of the hydrodynamic regime. This will allow us in Section \ref{sec:higher} to discriminate between the modifications to the gravitational effects due to dispersion and those due to the microscopic behaviour of the density.

\section{Vortices in trapped BECs: what is known}
\label{sec:known}

In an infinite condensate, the energy of a charge $\ell$ vortex is higher than the one of an array of $\ell$ singly charged vortices \cite{pethick2008bose}: this means that multiply quantized vortices are {\em energetically unstable}.

Concerning harmonically trapped condensates, extensive studies within the Bogoliubov approach~\cite{dodd1997excitation,rokhsar1997vortex} have shown that vortices -- even singly quantized ones -- are always {\em energetically unstable} since they possess a negative-energy $m=1$ mode localized around the vortex core, corresponding to precession around the trap center. Actual spiraling of the vortex out of the condensate requires some mechanism to dissipate the extra energy, for example via interaction with thermal atoms at finite temperatures~\cite{fedichev1999dissipative}. As a result, the vortex position remains {\em dynamically stable} under the purely conservative dynamics (\ref{eq:bogodegennes}).

\begin{figure}[t]
  \centering
  \includegraphics[width=\columnwidth]{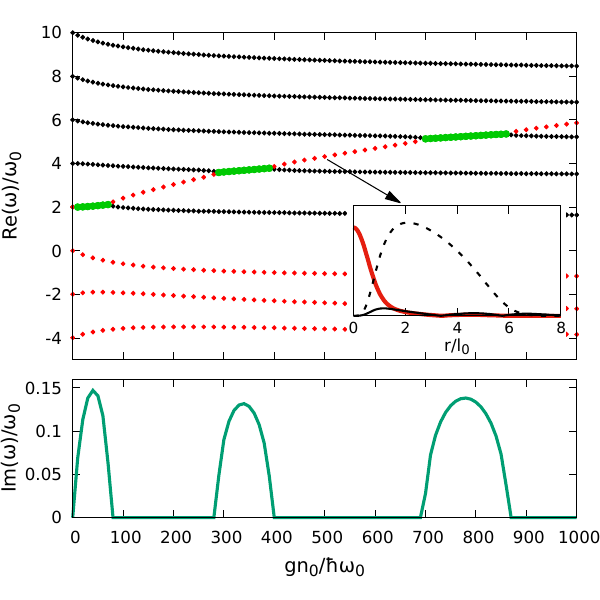}
  \caption{Real (upper plot) and imaginary (lower plot) parts of the BdG eigenfrequencies for modes of azimuthal number $m=2$ on a charge $\ell=2$ vortex in an harmonic trap for different values of the interparticle interaction energy. Here $n_0=N/l_0^2$ with $N$ number of atoms and $l_0=\sqrt{\h/(2M\w_0)}$. Black (solid), red (dotted) and green (thicker) lines correspond to positive-, negative- and zero-norm modes. In the inset a plot of the energetically unstable mode for $gn_0/\h\w_0=500$: the black (thinner) and red (thicker) lines show the modulus of the $u$ and $v$ components of the BdG spinor, the dashed line is a rescaled plot of the condensate density. }
  \label{fig:S2M2-harmonic}
\end{figure}

While all other excitation modes of singly charged vortices are energetically (and thus dynamically) stable, multiply quantized ones display alternate intervals of dynamical instability and stability as the nonlinear interparticle interaction is varied with respect to the trap frequency. An example of this behaviour for $m=2$ perturbations on a charge $\ell=2$ vortex, reproducing the conclusions of \cite{pu1999coherent}, is shown in Figure \ref{fig:S2M2-harmonic}. After numerically finding the radial profile $f(r)$ of the GPE ground state at fixed circulation $\ell$ with an external trapping potential $V_\mathrm{ext}(r)=M\w_0^2r^2/2$, we diagonalize the corresponding BdG matrix \eqref{eq:BdGmatrix} for a fixed azimuthal number $m$ but different values of the nonlinear parameter of the GPE, i.e. for different values of the interparticle interaction constant or for different numbers of atoms in the trap.

One can see that the system has a negative energy mode (negative norm and positive frequency) for all values of the parameters and is thus energetically unstable. This energetically unstable mode is localized near the vortex core, as can be seen in the inset.
As discussed also in~\cite{lundh2006dynamic}, dynamically unstable modes can emerge from the crossing of this negative-norm band with a positive-norm one. This is a consequence of the symplectic~\cite{arnol2013mathematical} nature of the BdG problem and in the Hamiltonian systems language is known as the mixing of modes of opposite Krein norm. This gives rise to the characteristic mode sticking visible in the upper panel, associated to the instability bubbles that are visible in the lower panel.

\begin{figure*}[t]
  \centering
  \includegraphics[width=0.8\textwidth]{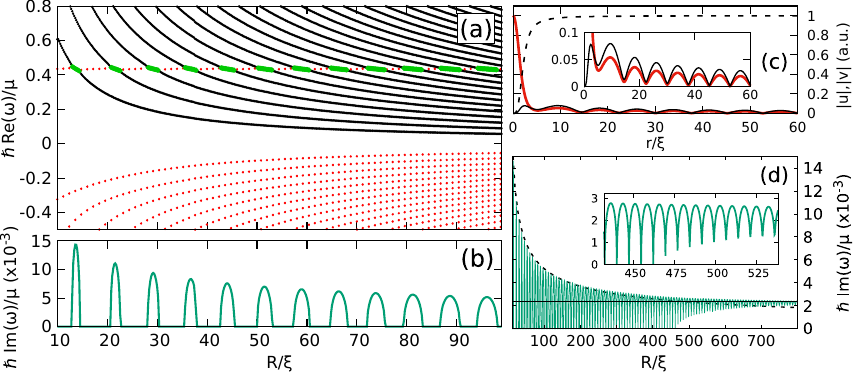}
  \caption{Real (a) and imaginary (b) parts of the Bogoliubov eigenfrequencies for modes of azimuthal number $m=2$ on a charge $\ell=2$ vortex in a BEC of size $R$. Black (solid), red (dotted), green (thicker) lines corresponds to positive-, negative-, and zero-norm modes. A wider view of the imaginary part is given in panel (d). Here, the $1/\sqrt{R}$ dashed line envelopes the instability maxima up to moderate $R$. The horizontal line indicates the instability rate extracted from the time-dependent simulation with absorbing boundary conditions shown in Figure \ref{fig:time-evol}.
  Panel (c): spatial shape of the  dynamically unstable core mode for the case with $R=60\xi$. The black (thin) and red (thick) lines respectively show the moduli of the $u_\phi$ and $v_\phi$ components of the Bogoliubov spinor. The dashed line shows the (rescaled) density profile of the vortex. In the inset the mode on a shorter amplitude range is plotted to highlight the structure at large $r$.}
  \label{fig:S2M2}
\end{figure*}

Even though the literature agrees on the occurrence of these instabilities and unambiguous experimental evidence is available~\cite{shin2004dynamical}, the situation is much less clear for what concerns the physical origin of the instability. The fundamental question that we plan to address in this article is to understand whether the dynamical instability is induced by the trapping, as some authors suggest~\cite{aranson1996stability,skryabin2000instabilities}, or whether it is instead an intrinsic instability of the vortex core, as recently argued in~\cite{takeuchi2018doubly}.

What is sure is that vortices of any charge in trapped condensates have some angular momentum channels with energetically unstable modes localized in the core. In the language of the gravitational analogy, these modes corresponds to negative-energy modes localized in the ergoregion of a rotating spacetime. The energetic instability can then be turned into a dynamical one if enough reflection happens on either side of the ergoregion: if it happens on the outer side, one has the so-called black-hole-bomb effect; if it happens on the inner side, an ergoregion instability occurs.

As it has been pointed out in \cite{takeuchi2018doubly} and we are going to show in the next sections, the instability of multiply charged vortices in condensates persists in spatially unbounded geometries where no reflection from the outer side can occur, so it can be classified of the ergoregion instability type. The effect of the trapping is rather the opposite, since it tends to suppress the instability within some specific regions of parameters as already visible in Figure \ref{fig:S2M2-harmonic}.

\section{A charge 2 vortex in an infinite BEC}\label{sec:charge-2}

\subsection{Large system limit $R\to \infty$}

As a first step, we follow the path of~\cite{takeuchi2018doubly} and investigate the stability of a doubly quantized vortex in an infinite BEC by looking how the spectrum of a finite system of size $R$ evolves in the infinite size limit $R\to \infty$. To this purpose, we numerically find the radial profile $f(r)$ of the GPE ground state $\Psi_\ell$ with a given circulation $\ell$ on a wide but finite interval $[0,R]$. In order to mimic a spatially homogeneous BEC, Neumann boundary conditions $\de_rf|_{r=R}=0$ are imposed on the BEC wavefunction. The BdG spectrum is then obtained imposing Dirichlet boundary conditions at $r=R$ onto the perturbation, $u_\phi(r=R)=v_\phi(r=R)=0$. The calculation is repeated for growing values of the size~$R$. 

The resulting discrete spectra of modes are shown in the left panels of Figure \ref{fig:S2M2} as a function of $R$ for an $m=2$ perturbation on a charge $\ell=2$ vortex. An energetically unstable mode (with negative norm and positive frequency) is clearly visible in panel (a) at an (almost) $R$-independent frequency around $0.44\,\mu/\hbar$. This $R$-independence is a strong indication that the mode is localized in the core region, which is further verified in the exponential decay of the envelope of $u_\phi$ and $v_\phi$ of the unstable mode (panel (c)). As for the trapped BEC case, dynamical instabilities emerge from the crossing of this negative norm core mode with the positive-norm collective modes; the resulting instability bubbles are shown in Figure \ref{fig:S2M2}(b). 

More insight in the instability mechanism is visible in the spatial profile of the dynamically unstable mode shown in panel (c): here, one recognizes a localized part at the vortex core which contributes with a negative norm and an extended part that penetrates deep in the bulk with a positive norm (better visible in the zoom in the inset). The two add up to a total zero norm, as expected for a dynamically unstable mode. This spatial structure indicates that the instability is due the coupling of a localized negative-energy excitation to a propagating positive-norm one. The oscillations that are visible at large $r$ are due to the interference of the outgoing waves with their reflection at the system edge $r=R$.

Looking at Figure \ref{fig:S2M2}(d), an important distinction between the moderate-$R$ and large-$R$ regimes jumps to the eyes. In the former case, the positive-norm collective modes of the condensate are well distinct in energy and the stability islands (instability peaks) are well separated as a function of $R$: stability (instability) occurs whenever the phase of the reflected waves at the $r=R$ boundary destructively (constructively) interferes with the oscillation of the core mode. This same mechanism is the cause of the finite stability windows that are also visible in the trapped condensate case of Figure \ref{fig:S2M2-harmonic}. A time-dependent insight on this destructive interference will be given in Section \ref{sec:superradiance}, where we discuss the temporal evolution of the system in response to a initial perturbation. In contrast, for large $R$ the instability bubbles merge with each other and the instability rate tends to a $R$-independent value.

Some qualitative understanding about this crossover can be obtained by analytical means. As mentioned in [11], the instability stems from the mixing of a spatially extended positive-norm mode with a localized negative-norm one, so the matrix element $\mathcal{M}$ of the mixing scales as the normalization of the spatially extended mode. In our cylindrical geometry, this normalization scales as $R^{-1/2}$, which determines the scaling $\mathcal{M}(R)\sim R^{-1/2}$ of the matrix element. This scaling reflects in an analogous scaling for the envelope of the instability rate maxima in the moderate-$R$ regime where modes are well discrete, see the dashed line in Figure \ref{fig:S2M2}(d).

The width of the instability bubbles is instead determined by the width of the regions around the crossing point for which the detuning of the positive- and negative-norm modes is not larger than the matrix element $\mathcal{M}$. The width in $R$ is hence proportional to the ratio between the matrix element and the derivative $d\omega/dR$, namely $\delta R\sim \mathcal{M}/|d\omega/dR|$.

Since the positive-norm modes have a phononic nature, the frequency of the $j$-th mode (with $j$ integer) scales approximately as $\omega_j\sim j\,c_s/R$, so that the derivative at the crossing point with the frequency $\omega_-$ of the trapped mode is given by $|d\omega_j/dR|\sim \omega_-/R$. Hence the width $\delta R$ of the instability bubbles increases as $\sqrt{R}$ for growing $R$, as visible in Fig.\ref{fig:S2M2}(b). Given the scaling of $\omega_j$ on $R$, the spacing $\Delta R$ between modes at the same frequency is instead a constant, so that neighboring instability bubbles eventually merge with each other into a broad continuum for large values of $R$.

More in detail, the spacing $\Delta R$ along $R$ can be related to the frequency spacing $\Delta \omega$ between modes at a given $R$ by $\Delta R = \Delta \omega / |d\omega/dR|$. Since the mode spacing $\Delta \omega$ is related by $\Delta \omega = 2 \pi/T_{\rm rt}$ to the round-trip time $T_{\rm rt}$ of phonons from the core to the $r=R$ boundary and back, the merging condition $\delta R \gtrsim \Delta R$ can be reformulated as $\mathcal{M}(R)\,T_{\rm rt}\gtrsim 1$ which has a transparent physical interpretation: the instability bubbles due to the spatially finite geometry disappear into a structureless continuum when the round-trip time exceeds the characteristic time of the instability, so that finite-size effects can no longer affect the dynamics of the instability.

\subsection{Outgoing boundary conditions}

\begin{figure}[t]
  \centering
  \includegraphics[width=\columnwidth]{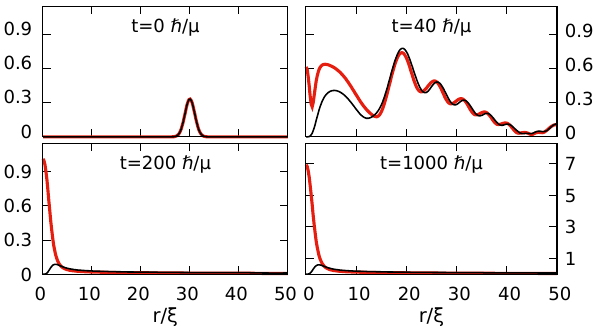}
  \caption{Snapshots of the time evolution of a $m=2$ perturbation scattering on a $\ell=2$ vortex. Black (thin) and red (thick) lines respectively show the $u_\phi$ and $v_\phi$ components of the Bogoliubov spinor. Outgoing boundary conditions are imposed by including a wide and smooth imaginary potential of Gaussian spatial shape centered at the edge of the integration box $r=700\xi$, of variance $120\xi$ and amplitude $0.15\mu$, so to effectively absorb the perturbation spinor $\ket{\phi}$ and suppress the reflected waves.}
  \label{fig:time-evol}
\end{figure}

While this way of taking the infinite-size limit may seem a sound way of describing a spatially infinite system, one must not forget that the eigenmodes of finite systems have a standing-wave shape and necessarily involve a reflected in-going wave, while for a truly infinite system no reflection is possible and the eigenmodes must have a purely outgoing character. As a consequence, the spectrum of a finite system is generally very different (even in the infinite-size limit) from the one in the asymptotic outgoing boundary conditions case. Because of this crucial difference, well highlighted for the Klein--Gordon case in~\cite{coutant2016dynamical}, it is thus essential to put any conclusion on the infinite system on solid grounds by implementing radiative boundary conditions where all reflected waves are removed from the outset.

In Figure \ref{fig:time-evol} we summarize a numerical study of the time-dependent BdG equations (\ref{eq:bogodegennes}) where such radiative boundary conditions are implemented by adding an effective absorption at large distances. A series of snapshots of the evolution of a perturbation in the $m=2$ channel in the presence of a $\ell=2$ vortex are shown, starting from a (arbitrarily chosen) real and Gaussian perturbation equally affecting the $u$ and $v$ components of the BdG spinor (upper-left panel). At early times (not shown) the perturbation splits in a pair of in- and out-going wavepackets the in-going one is then reflected by the vortex core and ends up propagating in the outwards direction as well, albeit with a strongly deformed shape because of the superluminal dispersion (upper-right panel). Eventually, the spatial shape of the perturbation is characterized at long times by an exponentially growing, negative-norm unstable core mode and a positive-norm excitation current propagating to infinity (bottom panels). Compared to Figure \ref{fig:S2M2}(c), the outgoing boundary conditions remove the interference-induced oscillations at large distance, leaving only the exponential spatial decay typical of unstable modes. The complex profile visible in the upper-right panel is a transient effect due to interference between the incident and the reflected wavepackets at the vortex core and disappears at late times as shown in the bottom row panels.

The temporal growth of the core mode can be precisely fitted with an exponential law (not shown) of instability rate $\Im(\omega) \simeq 0.00242\,\mu/\hbar$, indicated by the horizontal line in Figure \ref{fig:S2M2}(d) and in perfect agreement with the rate found for the finite system in the $R\to \infty$ limit. As a conclusion of this Section, our rigorous way of directly dealing with an infinite system perfectly confirms the results of \cite{takeuchi2018doubly} and offers further understanding of the validity of their infinite-size-limit procedure.

\section{Higher charge vortices}
\label{sec:higher}

\begin{figure}[t]
 	\centering
	\includegraphics[width=\columnwidth]{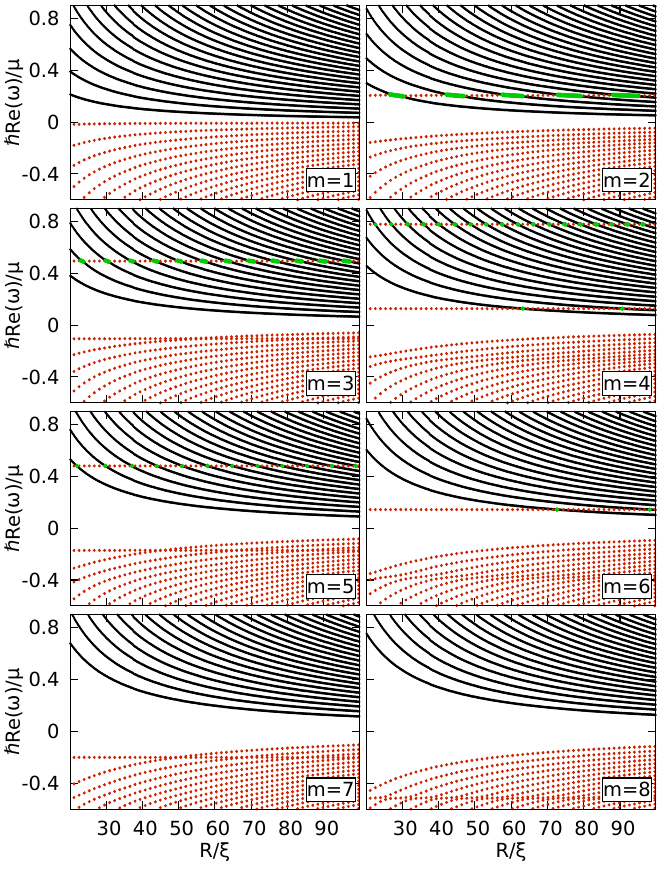}
	\caption{$R$-dependence of the Bogoliubov spectrum of a $\ell=4$ vortex for different azimuthal $m$. Black (solid), red (dotted) and green (thicker) lines correspond to positive-, negative- and zero-norm modes.}
	\label{fig:S4-spectra}
\end{figure}

Based on this important general result on the instabilities of a spatially infinite system, we are now entitled to apply the numerically simpler $R\to \infty$ infinite-size-limit procedure to more general cases, starting from $\ell>2$ charge vortices in uniform condensates. As an example, we display in Figure \ref{fig:S4-spectra} the $R$-dependence of the different-$m$ spectra for a given charge $\ell=4$. Independently from $\ell$, the $m=1$ spectrum always shows a negative-norm mode around zero energy. This core mode corresponds to the zero energy mode found in~\cite{pitaevskii1961vortex} and associated to the translation of the vortex core. For large but finite $R$, the frequency of this $m=1$ negative-norm core mode is very small and negative, meaning that the system is energetically stable. As expected, this frequency tends to $0$ in the $R\to \infty$ limit where translational invariance is recovered.

While the $m=1$ mode is the only core mode for $\ell=1$ vortices, for larger $\ell\geq 2$ other negative norm core modes appear for increasing $m$ at both negative and positive BdG frequencies, corresponding thus to positive and negative energies. Interestingly, the energy of the lowest energy (highest frequency) core mode decreases until $m=\ell$ and then starts increasing again until it becomes positive and energetic stability is recovered for all $m>2\ell-2$.
Since dynamical instabilities result from negative norm modes crossing the positive norm ones, this means that instabilities can only occur in the finite range of $m$ values
\begin{equation}
	2\le m\le 2\ell-2.
\end{equation}
Notice that this result differs from what is reported in previous literature: Ref.~\cite{pu1999coherent} reports dynamical instabilities only for $m\leq \ell$, while Ref.~\cite{lundh2006dynamic} claims that they exist for all $m\ge2$.

For what concerns the rate of the different instability channels, we find the quite unexpected result that for all values of $\ell$ the rate is strongest for $m=2$ and then decreases with $m$. Since the instability of multiply charged vortices is associated to their splitting into an assembly of $\ell$ singly charged vortices, one may have expected the most unstable mode to be at $m=\ell$. However our analysis shows that the vortex decay begins with lower-$m$ deformations of the core and the splitting in $\ell$ parts only appears during the later dynamics dominated by nonlinear effects.

\begin{figure}[t]
  \centering
  \includegraphics[width=\columnwidth]{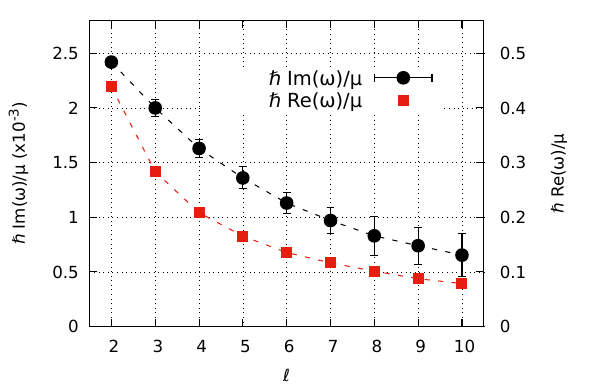}
  \caption{Imaginary (black circles) and real (red squares) parts of the $m=2$ unstable mode frequency for growing vortex charge $\ell$. The meaning of the error bars on the black circles is explained in the text.}
  \label{fig:charge-dependence}
\end{figure}

To investigate instead the $\ell$ dependence of the instability, we focus on the (most unstable) $m=2$ channel and for each $\ell$ we perform the infinite-size-limit procedure until the instability bubbles have fully merged into the large-$R$ unstructured continuum illustrated in Figure \ref{fig:S2M2}(d). To further remove numerical artifacts, we take the average of the last few $R$-dependent oscillations as the instability rate and the corresponding variance as an estimate of the error on this value. The result for vortices up to charge $\ell=10$ is shown in Figure \ref{fig:charge-dependence}, where also the real part of the frequency of the unstable modes is shown. Surprisingly the instability rate is highest for the lowest $\ell=2$ charges and decreases continuously while increasing $\ell$. This implies that the first stages of the vortex splitting process are slower for higher charge vortices.

It is now interesting to compare our results to the recent work~\cite{oliveira2014ergoregion} carried out for a purely hydrodynamic system for which the gravitational analogy holds exactly. With the aim of studying ergoregion instabilities, this work considers the case of a vortex without a drain and, since the hydrodynamic description breaks down near $r=0$, the KG equation must be supplemented with a reflecting boundary condition at a finite radius. For a given size of the ergoregion, the authors then find that all the high-enough $m$ modes are dynamically unstable, but the instability rate is stronger for the lower-$m$ unstable modes. The fact that this overall hierarchy of the instability rates of the different $m$-channels is shared by the Bogoliubov and the hydrodynamic calculations confirms that the nature of the instability is indeed the same in the two cases.

The main difference between the two calculations lies in the upper bound on the $m$ instability range. Also this feature is however compatible with the interpretation of the instability in terms of an ergoregion instability: it can in fact be ascribed to the superluminal dispersion that pushes the energy of the high-wavevector excitations modes towards high energies. This physical interpretation was numerically confirmed by solving the linearized problem for the excitation modes in the long-wavelength approximation. To this purpose we applied the same diagonalization procedure used for the BdG problem to the corresponding KG problem, that is equation \eqref{eq:bdg-densphase} without the derivative term in the upper-right element. This removes the dispersive effects. As expected, in this case we find no upper bound on $m$ for the occurrence of instabilities, so that unstable modes are present also for $m>2\ell-2$.

The fact that the same conclusions were obtained in the hydrodynamic calculations for a spatially homogeneous density profile of~\cite{oliveira2014ergoregion} suggests that the specific profile of the density around the vortex core does not have any substantial effect on the $m$-dependence of the instability rates. On the other hand, the density depletion around the vortex core seems to play an important role in the $\ell$ dependence of the instability rate: an increase of the instability rates with the vortex circulation was in fact observed in the hydrodynamic case of~\cite{oliveira2014ergoregion}, in stark contrast with the decrease found in our calculations shown in Figure \ref{fig:charge-dependence}. Since the same decrease of the instability rate was observed in the corresponding KG calculations where the dispersive effects are not present but the density depletion in the core remains, this difference can not be ascribed to the superluminal dispersion of Bogoliubov excitations, but rather to the density depletion around the vortex core.

As a main conclusion of this and the previous Sections, our joint numerical and analytical analysis fully confirms that the instability of multiply quantized vortices in BECs is a dispersive version of the ergoregion instability of rotating acoustic spacetimes and has no relation with the black hole bomb effect.

\section{On the stability of singly charged vortices}\label{sec:charge1}

\begin{figure*}[t]
  \centering
  \includegraphics[width=0.7\textwidth]{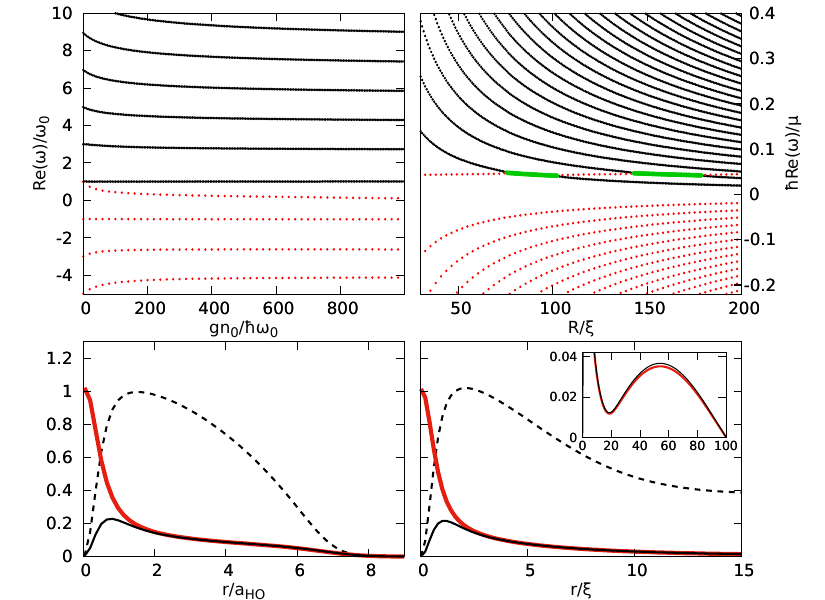}
  \caption{Left column: BdG spectra of a $\ell=1$ vortex in a harmonically trapped condensate as a function of the interparticle interaction energy (top).  Spatial profile of the energetically unstable mode for $gn_0=1000\h\w_0$ (thin black and thick red lines correspond to the moduli of the $u_\phi$ and $v_\phi$ components of the Bogoliubov spinor in arbitrary units) along with (dashed line) a rescaled version of the condensate density profile (bottom). Right column: BdG spectra of a $\ell=1$ vortex in the presence of an attractive Gaussian potential of the form \eqref{eq:Gaussian_V} of strength $A=2gn_\infty$ and spatial size $\sigma=5\xi$ (with $n_\infty$ asymptotic density and $\xi$ the associated healing length) as a function of the total radius $R$ of the condensate (top). Spatial profile of the dynamically unstable mode for $R=100\xi$ along with (dashed line) the condensate density profile (bottom).}
  \label{fig:vortex-bump}
\end{figure*}

In the previous Section we saw how, for all the vortex charges, the $m=1$ channel shows a core mode at zero frequency in the infinite system corresponding to a translation of the core: thanks to translational invariance, this displacement does not alter the energy of the system. Based on ergoregion instability arguments, the stability of charge $\ell=1$ vortices can be attributed to the dispersive effects that do not allow the presence of localized negative-energy modes at higher $m>1$.

In contrast to what is often claimed, the stability (both energetic and dynamical) of $\ell=1$ vortices in trapped geometries is not a general fact. For example, as we already said, vortex translation is energetically unstable in harmonically trapped BEC~\cite{dodd1997excitation}, whose inverted-parabola-shaped density profile favours expulsion of the vortex. The spectra for varying nonlinear interaction is visible in the top-left part of Figure \ref{fig:vortex-bump}, where one can see that a negative-energy core mode (whose spinor components are plotted in the lower panel) approaches zero frequency from above while increasing the number of atoms. The fact that this energetically unstable negative-norm mode is always located in frequency below the lowest positive-norm collective mode guarantees that the energetic instability never becomes dynamical.

Quite unexpectedly, if the density profile shows instead a bump surrounded by a wide region of lower density, this ordering of the modes is no longer guaranteed and collective modes satisfying the resonance condition for dynamical instability may be available for $m=1$, leading to a \textit{dynamical} instability of singly-charged vortices. This feature is illustrated in the right part of Figure \ref{fig:vortex-bump}, where the infinite-size-limit procedure is applied to a condensate subjected to an inverted-Gaussian-shaped potential
\begin{equation}
 V_\mathrm{ext}=-A\exp[-r^2/(2\sigma^2)]
 \label{eq:Gaussian_V}
\end{equation}
with a positive chemical potential $\mu>0$. The corresponding density profile is plotted with a dashed line in the lower panel: it displays a bump at short distances $r\lesssim \sigma$, then it tends to a non-vanishing constant value for large $r\gg \sigma$. In this configuration, instability regions appear as shown by the green lines in the top-right panel. In this parameter regions, the vortex starts precessing around the center with an ever-increasing radius, while its periodic rotatory motion leads to the emission of outward propagating sound waves in the external flat region. As usual, the positive energy of the emitted sound compensates for the negative energy associated to the vortex motion. In the harmonically trapped condensate, this sound emission process would not be possible as the cut-off frequency for collective modes in the condensate lies above the precession frequency of the vortex.

Further insight on this physics is obtained by comparing the spatial shape of the $m=1$ core modes plotted in the bottom panels for the two cases. This graphs show how these modes share the same shape near the core but, while the one in the harmonic trapped BEC quickly decays to zero as it has no collective mode to couple with (bottom-left panel), the one in the Gaussian potential couples to a long-wavelength collective mode that extends throughout the whole condensate, as it can be seen from the non-vanishing weight at large $r$ (bottom-right panel). 

As a final point, it is interesting to note in the inset of this plot how the $u_\phi$ part of the spinor slightly exceeds the $v_\phi$ one at large $r$. This is in contrast to the harmonically trapped case, in which $u_\phi$ and $v_\phi$ simultaneously vanish when the density approaches zero, and shows that the long-distance part of the excitation indeed contributes with a positive energy and norm and confirms that the instability results from the mixing of a localized negative-norm mode with an outward-propagating positive-norm mode.

\section{A time-dependent perspective: superradiant scattering and destructive interference}\label{sec:superradiance}

In the Introduction we mentioned the important fact that energetic instabilities can lead to superradiant scattering: the presence of negative-energy modes localized in the ergoregion makes it possible for incident positive-energy waves to be amplified during the reflection process at the expense of an equivalent negative energy being stored into the energetically unstable modes. Energy conservation is expressed here by the conservation of the Bogoliubov norm \eqref{eq:bogo-norm}.

Dynamical instabilities emerge if the trapped negative-energy excitations remain in the system and get further amplified in a stimulated way while emitting positive-energy excitations outside the ergoregion. Superradiant scattering from a dynamically stable configuration occurs instead if the negative-energy modes are quickly removed from the system and are no longer available to stimulation: this can happen, for example, when an horizon is present inside the ergoregion and the energy in the localized mode rapidly falls behind it.

While the presence of such an \textit{absorption} mechanism inside the ergoregion has often been regarded as a necessary condition to have superradiance (see for example \cite{richartz2009generalized}), this is not generally the case: amplified scattering just depends on the presence of an ergosurface and not on the boundary conditions of the problem, that may be specified even very far from the ergosurface itself. While a complete discussion of this physics will be given in our forthcoming work \cite{giacomellipreparation}, here we restrict ourself to a quick numerical overview of this phenomenon in the present case of dynamically unstable multiply quantized vortices.

\begin{figure}[t]
 	\centering
	\includegraphics[width=0.95\columnwidth]{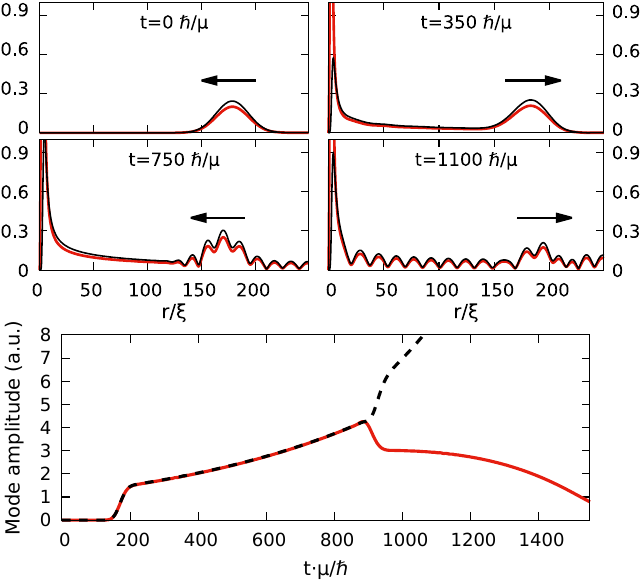}
	\caption{First and second row: Snapshots of the evolution of a Gaussian Bogoliubov wavepacket of $m=2$ incident waves hitting a charge $\ell=4$ vortex. The frequency of the incident wavepacket is centered around $\omega=0.2\h/\mu$, namely the frequency of the core mode visible in the spectra shown on the top-left panel of Figure \ref{fig:S4-spectra}. Dirichlet boundary conditions are imposed to the Bogoliubov modes at $R=380\xi$. The arrows indicate the direction of the radial group velocity. The bottom panel shows the time-dependence of the core mode amplitude (red solid line), as measured by the $r=2\xi$ value of the $v_\phi$ component of the BdG spinor. The black dashed line is the time-dependence of the same core mode amplitude for a slightly different size $R=388.4\xi$, for which the system is dynamically unstable. Around $t=900 \mu/\hbar$, the two curves show clear signature of the destructive vs. constructive interference between the reflected wavepacket and the core mode.}
	\label{fig:scattering-evol}
\end{figure}

Superradiant amplified scattering is illustrated in Figure \ref{fig:scattering-evol} where we show snapshots of the evolution of a Gaussian wavepacket of azimuthal number $m=2$ incident onto a $\ell=4$ vortex in a spatially homogeneous condensate. In practice, we have numerically obtained the condensate profile as done in Section \ref{sec:charge-2}. We have then constructed a wavepacket of Bogoliubov excitations centered at the frequency of the core mode found in Figure \ref{fig:S4-spectra} and with a group velocity directed towards the vortex core.

The wavepacket reaches the vortex core between the first ($t=0$) and the second ($t=350 \mu/\hbar$) snapshots. While being reflected from it, it populates the (negative energy) core mode as visible in the narrow excitation peak on the left edge of the second to fourth panels. Even though pulse distortion effects make it hardly visible by eye, a quantitative comparison of the BdG norm \eqref{eq:bogo-norm} of the wavepacket before and after the scattering process shows that the reflected packet is amplified by approximately $18\%$. This is a clear evidence of the occurrence of superradiant scattering in the short-time dynamics.

At later times, after being triggered by the superradiant scattering process, the amplitude of the core mode keeps exponentially increasing between $t=200 \h/\mu$ and $t=800 \h/\mu$, as it is visible in the time-dependence shown in the bottom panel. Of course, this increasing amplitude is associated to an analogous exponential growth of the excitations in the bulk behind the amplified wavepacket.

While this first stage of the evolution does not depend on the actual size of the system and is well captured by the theory for an infinite condensate, the long-time dynamics crucially depends on the specific value of the system size $R$. In the simulation of Figure \ref{fig:scattering-evol}, this was chosen to fall within one of the stability windows. In the third snapshot at $t=750 \hbar/\mu$, the amplified packet has been already reflected by the Dirichlet boundary condition at the end of the system at $R=380\, \xi$ (outside the plot) and is approaching again the vortex core. The complex oscillatory profile visible in this third snapshot is due to the interference of the reflected wavepacket with the exponentially growing emission by the unstable core mode. A second scattering process occurs when this reflected wavepacket hits the vortex: during this scattering process, the incident wavepacket interferes with the excitation amplitude left in the core mode after the first scattering and, since then, exponentially growing. For the chosen value of the condensate radius $R$, destructive interference occurs and the amplitude of the reflected wavepacket is suppressed, as shown in the last snapshot at $t =1100 \hbar / \mu$.

The complete time dependence of the negative norm core mode amplitude is shown in the bottom panel. As a signature of the destructive interference effect (solid red line), the exponential increase of the core mode amplitude suddenly stops around $t=900\h/\mu$ and is replaced by a strong decrease at later times.
After this first back-and-forth motion, the core mode evolution keeps displaying alternate intervals of increasing and decreasing amplitude. 

This complex time-dependent behaviour reflects the
dynamical stability of the system as predicted by the
time-independent calculation of the spectrum and is a
clear illustration of how the finite size of the system is
able to quench the ergoregion instability that would instead appear in
an infinite geometry. In agreement with the results presented in Sec.IV
for the large $R$ behaviour, the interference effect ceases being
relevant for $\gamma(R) T_{\rm rt}\gtrsim 1$, that is when the
round-trip time $T_{\rm rt}$ is so large that the core mode amplitude
has time to grow to such large values that the reflected wavepacket no
longer has any significant effect on the exponential growth of the
instability.

In the dynamically unstable regions, the interference at each scattering event would instead be constructive, leading to a ever increasing core mode amplitude with an even higher rate than in the infinite system. The effect of this constructive interference on the core mode amplitude for a slightly different system size $R=388.4\xi$ is plotted in the lower panel of Figure \ref{fig:scattering-evol}. As expected, in this case (black dashed line) the constructive interference leads to a sudden upwards jump of the core mode amplitude, which then keeps exponentially increasing.

\section{Conclusions}
\label{sec:conclu}

In this work we have made use of the analogy with rotating spacetimes to investigate the physical origin of the instabilities of multiply charged vortices in two-dimensional Bose--Einstein condensates. As it happens in the ergoregions of gravitational systems, the instabilities can be directly associated to the presence of energetically unstable modes in the vortex core region where the flow is supersonic, so they can be classified of an ergoregion nature. The differences with gravitational systems -- in particular the suppression of instabilities at high values of the angular momentum -- can be ascribed to the superluminal dispersion of Bogoliubov waves in condensates.

In stark contrast to black-hole bomb instabilities triggered by reflections outside the ergoregion, here the finite size of the system rather leads to a quench of the instability via a destructive interference effect. This explains the peculiar stability properties of vortices in trapped condensates and also proves that the dynamical instabilities are not, as it is sometimes claimed, induced by the trapping. Based on our novel understanding of these instability phenomena, exotic geometries where singly quantized vortices are also dynamically unstable were identified. 

From the analogue gravity point of view, our calculations show the robustness of superradiant phenomena against the superluminal corrections to the linear sonic dispersion and give an example of how the gravitational analogy can provide qualitatively correct results even outside its hydrodynamical regime of validity. Time-dependent calculations show how the presence of dynamical instabilities does not prevent the possibility of amplified scattering processes and proves that an horizon (or another dissipation mechanism inside the ergoregion) is not necessary for superradiant scattering to take place. 

Beyond the cylindrical vortex geometry considered here, a forthcoming publication~\cite{giacomellipreparation} will investigate these effects in the context of the Schiff-Snyder--Weinberg effect, a process that can be understood as an ergoregion instability in a different geometry. In particular, we will look at translationally invariant geometries where the different microscopic mechanisms of superradiance and dynamical instabilities can be identified and isolated. Well beyond the most popular case of atomic Bose-Einstein condensates on which we have focused our presentation, our conclusions are directly applicable to generic quantum fluids showing quantized vortex excitations, in particular quantum fluids of light~\cite{carusotto2013quantum,vocke2018rotating,fontaine2018observation}, where such phenomena are presently under active study~\cite{jacquet2020fluids} .

\bibliographystyle{apsrev4-1}
\bibliography{biblio.bib}

\end{document}